\documentclass[]{aa}
\usepackage{graphicx}
\usepackage{color}
\usepackage{epstopdf}
\usepackage{txfonts}
\usepackage{hyperref}
\usepackage{fixltx2e}
\usepackage{dblfloatfix}
\usepackage{float}
\usepackage[round]{natbib}
\begin{document}

\title{Rieger-type periodicity in the total irradiance of the Sun as a star during solar cycles 23-24}

\author{Gurgenashvili, E. \inst{1,2,3}, Zaqarashvili, T.V.\inst{4,2,3},  Kukhianidze, V. \inst{2,3}, Reiners, A. \inst{1}, Oliver, R. \inst{6}, Lanza, A.F. \inst{5}, Reinhold, T.\inst{7}}
\institute{
            %School of Natural Sciences and Medicine, Ilia State University, Cholokashvili ave. 3/5,  Tbilisi, Georgia\\
     %    \and
      %     E.Kharadze Georgian National Astrophysical Observatory, Mount Kanobili, Georgia \\
    %     \and
           %Institute of Physics, IGAM, University of Graz, Universit\"atsplatz 5, 8010 Graz, Austria \\
     %   \and
           Georg-August-Universit\"at, Institut f\"ur Astrophysik, Friedrich-Hund-Platz 1, 37077, G\"ottingen, Germany \\
           \and
           School of Natural Sciences and Medicine, Ilia State University, Cholokashvili ave. 3/5,  Tbilisi, Georgia\\
           \and
           E.Kharadze Georgian National Astrophysical Observatory, Mount Kanobili, Georgia \\
           \and
           Institute of Physics, IGAM, University of Graz, Universit\"atsplatz 5, 8010 Graz, Austria \\
           \and
        INAF-Osservatorio Astrofisico di Catania, Via S. Sofia, 78 - 95123 Catania, Italy \\
       \and
      Departament de F\'isica \& Institut d'Aplicacions Computacionals de Codi Comunitari (IAC3), Universitat de les Illes Balears, 07122 Palma de Mallorca, Spain\\
      \and
      Max-Planck-Institut f\"ur Sonnensystemforschung, Justus-von-Liebig-Weg 3, 37077 G\"ottingen, Germany \\
}

%\date{Received 30 December 20016 / Accepted 25 May 2017}
\abstract{Total solar irradiance allows for the use of the Sun as a star for studying observations of stellar light curves from recent space missions.}{We aim to study how the mid-range periodicity observed in solar activity indices influences the total solar irradiance.}{We studied periodic variations of total solar irradiance based on SATIRE-S and SOHO/VIRGO data during solar cycles 23-24 on timescales of Rieger-type periodicity. Then we compared the power spectrum of oscillations in the total solar irradiance to those of sunspot and faculae data  to determine their contributions.}{Wavelet  analyses of TSI data reveal strong peaks at 180 days and 115 days in cycle 23, while cycle 24 showed  periods of 170 days and 145 days. There are several periods in the sunspot and faculae data that are not seen in total solar irradiance as they probably cancel each other out through simultaneous brightening (in faculae) and darkening (in sunspots). Rieger-type periodicity is probably caused by magneto-Rossby waves in the internal dynamo layer,  where the solar cyclic magnetic field is generated. Therefore, the observed periods in the total solar irradiance and the wave dispersion relation allow us to estimate the dynamo magnetic field strength as 10-15 kG.} {Total solar irradiance can be used to estimate the magnetic field strength in the dynamo layer. This tool can be of importance in estimating the dynamo magnetic field strength of solar-like stars using light curves obtained by space missions.}
%
%\keywords{}

\titlerunning{.}

\authorrunning{Gurgenashvili et al.}

\maketitle

%=====================
\section{Introduction}
%=====================

The recent NASA space missions Kepler \citep{Borucki2010} and  Transiting Exoplanet Survey Satellite, (TESS, \citet{Ricker2014}) %\LEt{ Please merge parentheses here and in the following.}
, along with ESA's   Convection, Rotation and planetary Transits mission (CoRoT, \citet{Baglin2008}) collected a huge store of information on the basic properties of stars and exoplanets. These missions have provided the light curves for hundreds of thousands of stars, which allows us to gain new insights into stellar interiors using asteroseismology, as well as into stellar activity based on the time modulation of the curves. Stellar light curves may reveal stellar rotation periods \citep{Nielsen2013, Reinhold2013, McQuillan2014} and short-term activity cycles \citep{Ferreira2015, Reinhold2017}. To understand stellar rotation and activity, it is of fundamental importance to use the Sun as a star analog since it provides much more detailed information on solar rotation and activity variations.

\begin{figure*}
\includegraphics[width=18cm]{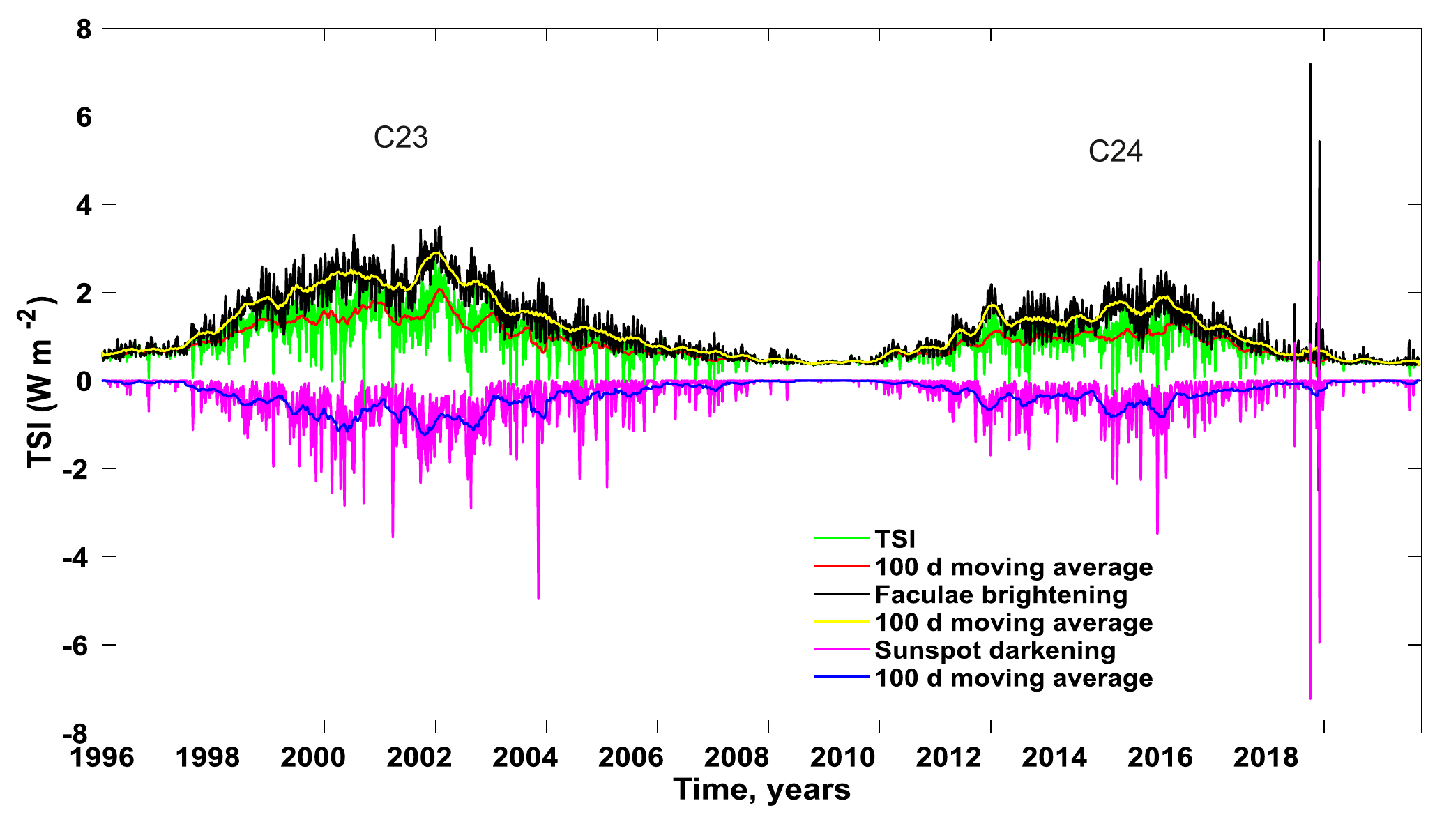}
\caption{Total solar irradiance (green line) during solar cycles 23-24 from SATIRE-S data in units of W m$^{-2}$. Red curve shows 100-day moving average of TSI. Black line corresponds to the faculae brightening (with yellow curve as the 100 day average), while the purple line shows the contribution of the sunspot darkening (blue curve as the 100-day average).}
 \label{fig:1}
\end{figure*}

The analog for stellar light curves is the total solar irradiance (TSI). The variability of  solar irradiation was first found after the launch of the Nimbus 7 mission in 1978. Subsequently, radiometers on board different satellites, such as Solar Maximum Mission (SMM) ACRIM-I,  NOAA-9, NOAA-10, SOHO/VIRGO, and SORCE/TIM  continuously measured the solar bolometric flux \citep{Hathaway2010}. The variations in the proxies for solar activity (sunspot number; total sunspot area) from the Maunder minimum up through today has been reconstructed using different methods \citep{Krivova2009, Krivova2010, Yeo2014,Yeo2017, Dasi-Espuig2016, Wu2018}.
Using solar full-disc magnetograms and continuum images,  \citet{Yeo2014} reconstructed the daily variation of the solar irradiance from 1974 to 2013, using the  Spectral And Total Irradiance REconstructions model (SATIRE).

The irradiance observations show that solar brightness changes over different timescales, ranging from several minutes to years. The temporal variation of the TSI is mainly caused by solar magnetic field effects, including sunspot darkening and faculae brightening.

It is important to study the relation between sunspots and faculae, as it has the potential to deliver vital knowledge of how the TSI changes, which can be used to better understand the stellar light curves observed by Kepler.
Sunspots and faculae are the most significant contributors to solar activity. Sunspots appear as dark spots on the solar surface, where the temperature is significantly lower. In contrast, faculae are brightest structures where the temperature is higher and they can be easily observed at the solar limb and almost invisible in the center of the disk. At the maximum of the solar cycle, the contribution from faculae dominates the sunspot contribution, making the Sun appear brightest at the maximum.

The Rieger-type periodicity of 150-200 days was discovered by Rieger et al. (1984) more than 30 years ago. This periodicity is seen in many indices of solar activity  \citep{Bai1987, Bai1990, Lean1990, Ballester2002, Ballester1999, Oliver1998, zaqarashvili2010}. \citet{Lean1989} showed that the Rieger period of 155 days is seen in the sunspot blocking function, 10.7 cm radio flux, and the Z\"urich sunspot number during solar cycles 19-21, but it is absent in plages, where the magnetic field is weak. This may indicate that the periodicity is related to the solar interior, namely with  the solar tachocline, where the solar magnetic field is thought to be generated. \citet{zaqarashvili2010} suggested that the periodicity can be explained by unstable magnetic Rossby waves in the tachocline owing to the latitudinal differential rotation and the toroidal magnetic field, which may lead to the quasi-periodic eruption of magnetic flux towards the surface. It has also been shown that the Rieger-type periodicity is anti-correlated with the solar cycle strength, such that stronger cycles exhibit shorter Rieger periods \citep{gurgenashvili2016, gurgenashvili2017}. Therefore, the observed periodicity and dispersion relation of magnetic Rossby waves can be used to estimate the magnetic field strength in the dynamo layer \citep{zaqarashvili2018}. Rieger-type periodicity can also be observed in the activity of other stars; therefore, the method may allow for the probing of the dynamo magnetic field on the stars at different stages of evolution.

In order to observe Rieger-type periods on other stars, we can look into light curves provided by Kepler, CoRoT and TESS. The existence of short-term cycles, which are similar to the Rieger periods, was previously reported based on the CoRoT \citep{Lanza2009}  and Kepler \citep{Bonomo2012,Lanza2019} observations. However, more observations and detailed analysis of stellar light curves are needed to improve on these results. Prior to a stellar light curve analysis, it is clearly of vital importance to study the periodicity in the total irradiance of the Sun in detail. In this paper, we analyze the TSI data for cycle 23 and 24 and we look for quasi-periodic variations on timescales of  Rieger-type periodicities, using the Morlet wavelet transform \citep{Torrence1998}.

\begin{figure*}
\includegraphics[width=18cm]{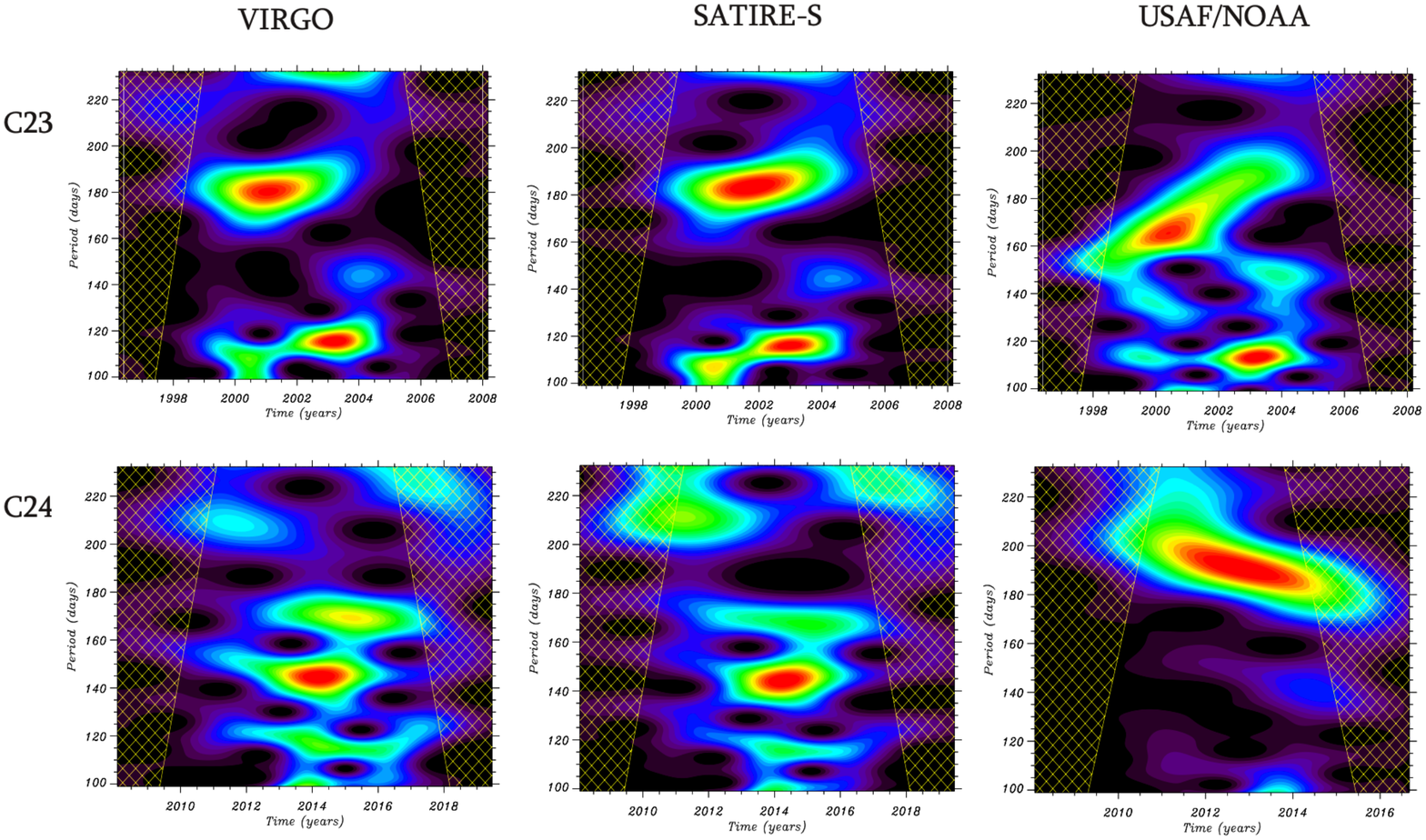}
\caption{Morlet wavelet analysis in the period range of  100-220 days \citep{Torrence1998}. Upper and lower panels correspond to the cycle 23 and 24, respectively. Left panels show the wavelet spectrum based on VIRGO data. Middle panels correspond to the SATIRE data. Right panels display the USAF/NOAA full disk sunspot area data.}
 \label{fig:7}
\end{figure*}

%=====================
\section{Data}
%=====================
We used three different data sets, including both observations and reconstructions. For the TSI measurements, we used data from Variability of solar IRradiance and Gravity Oscillations (VIRGO) on board Solar and Heliospheric Observatory (SOHO), combining two radiometers (Diarad and PMO6-V) and two sun-photometers, which measure the spectral irradiance at 402, 500, and 862 nm with 5 nm passbands.
To make comparisons with the SOHO/VIRGO TSI data, we used the SATIRE-S reconstruction model (here "S" stands for the Satellite era) from the Max Planck web page %\LEt{ set link as footnote.}
 \footnote{\url{http://www2.mps.mpg.de/projects/sun-climate/data.html}.}, which includes the data of the TSI and Spectral Solar Irradiance (SSI) starting from 1947 \citep{Yeo2014}.
It combines the contributions of faculae brightening and sunspot darkening.

The VIRGO data has much higher quality than the other older radiometers, however, an almost seven-month gap in the data appears during the ascending phase of cycle 23 \citep{Frohlich1995}. In order to search periodicities in VIRGO data, we filled the gap with the SATIRE-S reconstruction. For sunspot data, we use  the U.S. Air Force/National Oceanic and Atmospheric Administration (USAF/NOAA) sunspot area (SA) data %\LEt{ footnote.}
\footnote{\url{http://solarscience.msfc.nasa.gov/greenwch.shtml}.}, that has continuous data for cycles 23-24, without any gaps; however, the data ends in September 2016 and therefore does not cover the most recent years of cycle 24.

%=====================
\section{Results}
%=====================

Figure ~\ref{fig:1} shows TSI data using the SATIRE-S reconstruction. Total solar irradiance includes the contributions from different sources, mainly from faculae and sunspots. Faculae brightening (FB) leads to an increase in TSI, while sunspot darkening (SD) reduces it. Figure~\ref{fig:1} shows that the faculae contribution has almost twice as much amplitude as the sunspots in both cycles. Several different timescales are seen in the TSI. First of all, there is a clear modulation of TSI with the 11-year solar cycle.
In particular, TSI mainly follows the FB curve, which shows that the solar irradiance is faculae dominated. The second timescale is on the order of 0.5-1.5 years (from the range of the Rieger periodicity to that of the quasi-biennial oscillation) and the third timescale is on the order of solar rotation ($\sim$27 d). Here we are interested in searching for the mid-range periodicity in the total irradiance of the Sun.

\begin{figure*}
\includegraphics[width=16cm]{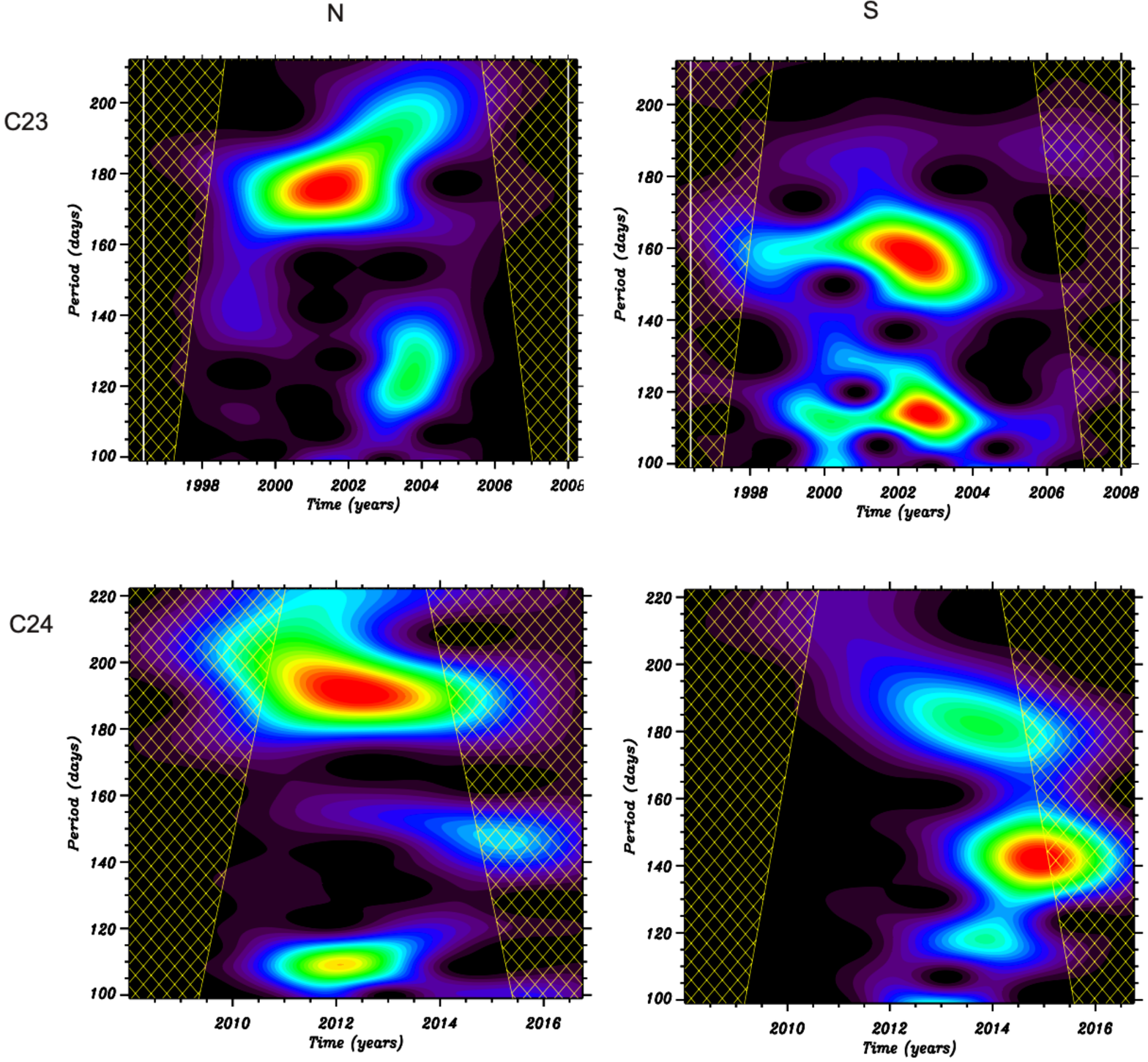}
\caption{Wavelet spectra of hemispheric sunspot area data during solar cycles 23 (upper panels) and 24 (lower panels) in the period range of Rieger periodicity (100-200 days). Left panels: Northern hemisphere, right panels: Southern hemisphere.}
 \label{fig:10}
\end{figure*}

Figure ~\ref{fig:7} shows the Morlet wavelet analysis of TSI and sunspot area during cycles 23-24 in 100-220 day period window (the periods are defined with close to 5-day accuracy). The hatched regions on the figure (and on all wavelet figures) show the cone of influence (COI), which means that the wavelet transform is not reliable in this areas \citep{Torrence1998}. First of all, we note that the VIRGO (left panel) and SATIRE (middle panel) data show almost identical spectra; therefore, SATIRE data can be safely used for detailed analysis.
The wavelet spectra are significantly different in the cycles 23 and 24; thus, we consider the periodicities in both cycles separately.

\subsection{Cycle 23}

 Two strong periodicities are seen in both the power spectra of the TSI data and the SATIRE-S model, namely: 180 days and 115 days. The peak at 180 days is in the range of typical Rieger-type periodicity (150-180 days) found previously in the sunspot area data \citep{gurgenashvili2016}. On the other hand, the peak at 115 days is relatively shorter compared to the typical Rieger periodicities, therefore it is probably related with another branch of periodicity (the shorter period can be explained by another harmonic of Rossby waves; see discussion in Section 4). There is also a weak power near the period of 140 days.
In order to compare the TSI variations with the sunspot area, we plot the wavelet spectrum of USAF/NOAA data on the right panel of this figure. A strong peak is clearly seen at 115 days, however, the strong power is absent at the period of 180 days. Instead, there is a strong peak near the period of 165 days. We can note the weak peak at 180 days, but it has much less power than the one shown at 165 days.
\citet{gurgenashvili2017} showed that cycles with substantial north-south asymmetry generally display two periodicities corresponding to hemispheric activity. The shorter periodicity is usually revealed in the more active hemisphere. The cycle 23 was south-dominated; therefore, the 165 day and 180 day periods may correspond to the southern and northern hemispheres, respectively.  The upper panels of Figure~\ref{fig:10} show the wavelet spectra of hemispheric sunspot areas in cycle 23. Indeed, north hemispheric data exhibit strong power at around 175 days, while the southern hemisphere shows a result near 160 days. Additionally, the northern hemisphere has a weak peak at 120-125 days, while the southern hemisphere has a strong peak at 115 days.
A comparison of the hemispheric sunspot wavelet spectrum (Figure ~\ref{fig:10}) to that of TSI (Figure ~\ref{fig:7}) shows that the long period in TSI (175-180 day) corresponds to the similar periodicity in the sunspot area of the northern hemisphere, while the shorter period in TSI (115 day) corresponds to the similar periodicity in the sunspot area of southern hemisphere.

\begin{figure*}
\includegraphics[width=18cm]{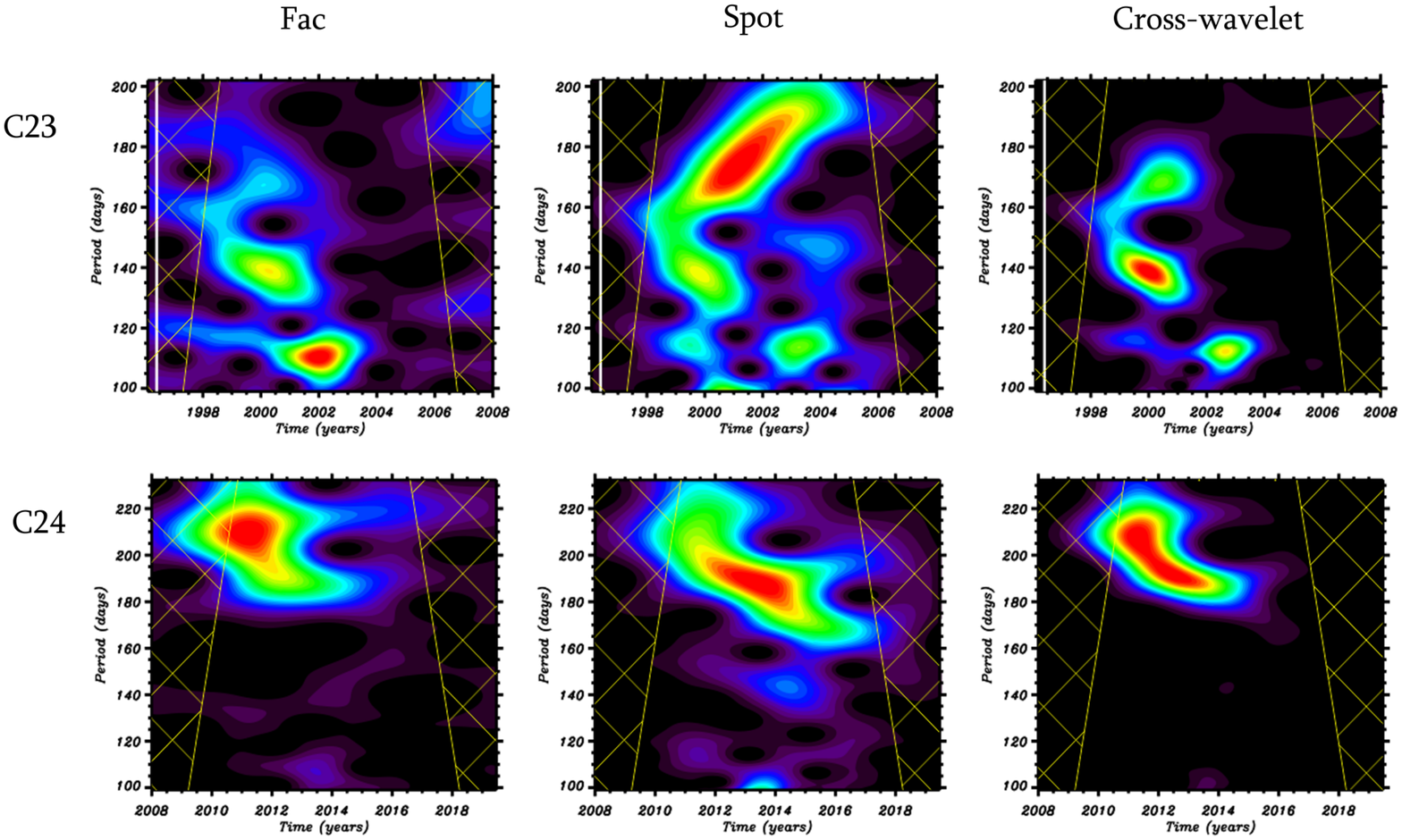}
\caption{Faculae brightening and sunspot darkening contributions in wavelet spectrum of TSI in the period range of Rieger periodicity (100-200 days) in solar cycles 23 (upper panels) and 24 (lower panels). Left and middle panels show the faculae and sunspot contributions respectively, while the right panels show the corresponding cross-wavelet powers.}
\label{fig:8}
\end{figure*}

\begin{figure}
\includegraphics[width=9cm]{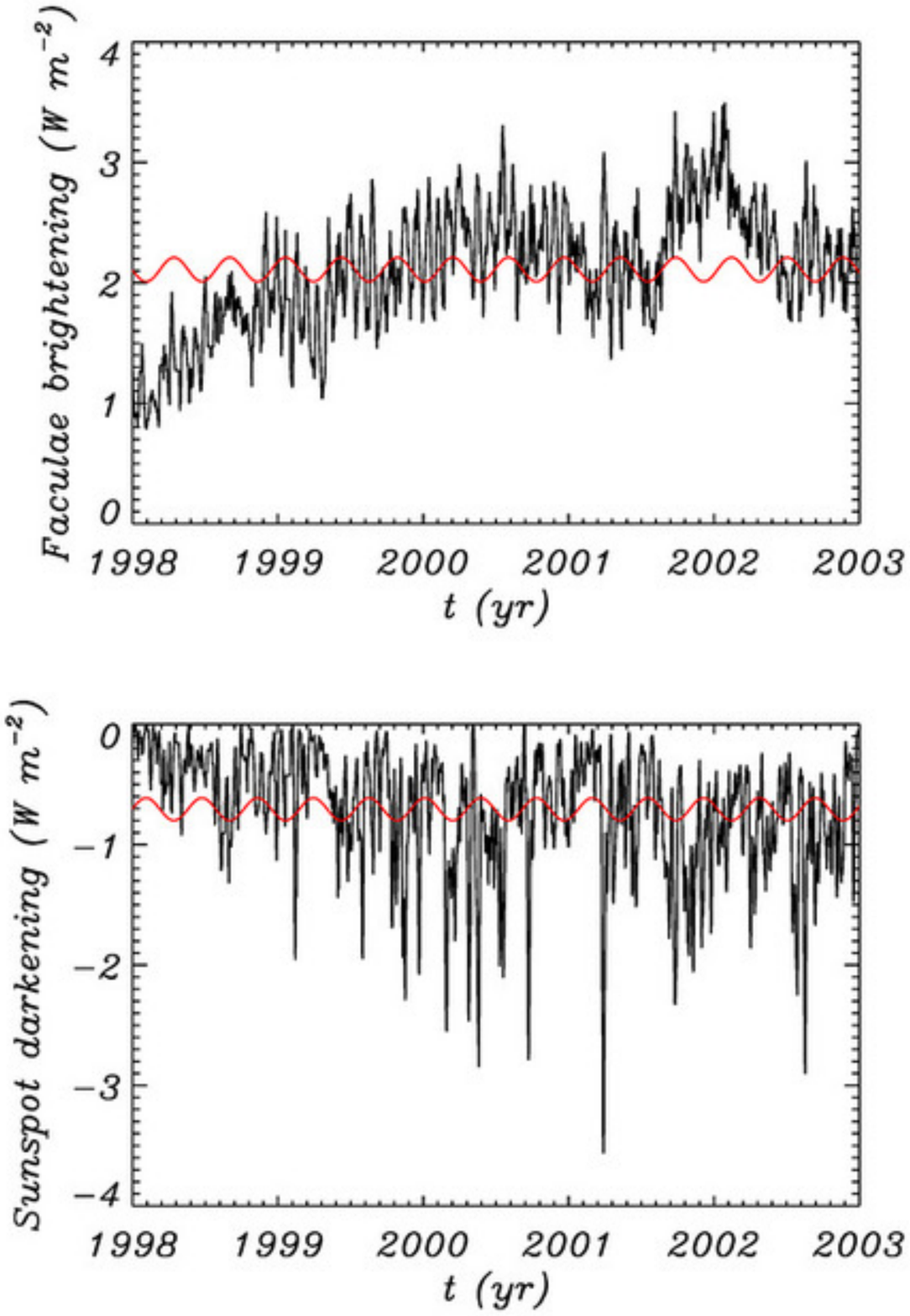}
\caption{Fitting of sinusoidal function with the period of 140 days (red curves) defined by Eq.~(\ref{eq:fit_funct}) to the raw data of faculae brightening (upper panel) and sunspot darkening (lower panel). We note that the sinusoidal fit here is only used to show the phase relation of the two oscillations and does not intend to model the time series.}
 \label{fig:11}
\end{figure}

The SATIRE-S reconstructed data allow us to study the contributions from faculae brightening and sunspot darkening in the TSI wavelet spectra. Figure ~\ref{fig:8} shows the wavelet spectra of the faculae brightening and sunspot darkening separately. Several interesting features can be seen in this figure. First, the strong periodicity of 180 days revealed in TSI corresponds to the sunspot contribution as it is not seen in faculae data. Second, another strong peak at 115 days in TSI spectrum is seen in both sunspot darkening and faculae brightening; therefore, it probably results from the joint action of faculae and sunspots. Finally, the two periods of 140 and 165 days, which appear in the faculae and sunspot data, are totally absent in the TSI spectrum. The possibility is that the oscillations cancel each other when TSI variations are considered.
%And the weak peak at the period of 220 days, which appear
To verify this explanation, we fit the following function both to the facular brightening and the sunspot darkening:

\begin{equation}\label{eq:fit_funct}
  A\cos\left(\frac{2\pi}{P}t+\phi\right) + C,
\end{equation}

\noindent where $A$ (in $W/m^2$), $P$ (in days) and $\phi$ (in radians) are the amplitude, period and phase of the sinusoidal component, respectively, while $C$ is a constant offset and $t$ is the time. The function fitting was done with the MPFIT library \citep{more1978}, using the IDL implementation by \citet{markwardt2009} %which is available at %\LEt{ footnote.}
\footnote{\url{http://purl.com/net/mpfit}}. A fitting is done for the raw data, with the following restrictions imposed on the fit parameters: $A>10^{-5}$, to avoid $A=0,$ which would result in just fitting a horizontal straight line to the data, $P$ fixed and $|\phi|<\pi$. For the 140~d periodicity, the fit was performed for the interval of 1998-2003, when the oscillations were seen in both the faculae brightening and the sunspot darkening. The results are: $A_{FB}=0.1004$, $\phi_{FB}=1.163$, $C_{FB}=2.111$ for the faculae brightening and $A_{SD}=0.0936$, $\phi_{SD}=-1.975$, $C_{SD}=-0.706$ for the sunspot darkening; $P$ is set to 140~d and is not allowed to vary. The result of the fit is shown in Figure ~\ref{fig:11}.

The phase difference, $|\phi_{FB}-\phi_{SD}|$, equals 3.138$\approx \pi$, hence the two oscillations are  in anti-phase during the interval and cancel out because their amplitudes are very similar. This explains why the 140~d periodicity is present simultaneously in the faculae brightening and the sunspot darkening, but is absent in the TSI. The same analysis has also been performed for the 165~d periodicity for the interval 1997-2003 and it was found that $|\phi_{FB}-\phi_{SD}|=2.535$. Therefore, the absence of this periodicity in the TSI cannot be explained by the cancelation in anti-phase of faculae brightening and sunspot darkening oscillations. We argue that this periodicity is not visible in the TSI wavelet diagram because of its low  power and the fact that it is masked by the much stronger 180-day sunspot darkening periodicity.

\begin{table}[h!]
\begin{tabular}{ |p{0.6cm}||p{0.8cm}|p{0.8cm}|p{0.6cm}| p{0.6cm}| p{0.6cm}| p{0.6cm}|  p{0.6cm}|}

 \hline

 Cycle & VIRGO  &SATIRE & fac & spot& SA &SA &SA  \\
    N  &        &       &     &     &      &N   &S \\

 \hline
 \hline
 C23   &  180   &180    &     &180 & 160  & 175 &160  \\
       &  115   &115    &115  &115 & 115  & 125 & 115  \\
 C24   &  170   &       &210  & 185& 190  & 190 &180\\
       &  145   &  145  &     &     &      & 110 & 115 \\
\hline

 \hline

\end{tabular}
\caption{Most significant periods (d) obtained from different data sets (VIRGO/SOHO, SATIRE-S, USAF/NOAA sunspot area (SA), faculae and sunspot contributions). The periods are defined up to a 5-day accuracy.}
\label{table:1}
\end{table}

We did a similar analysis for the 115 d periodicity during the interval of 1998-2006, when the oscillation was seen in both indicators and also in the TSI. The resulting parameters are $A_{FB}=0.0588$, $A_{SD}=0.0600m$ and $|\phi_{FB}-\phi_{SD}|=4.074$.
The values of the offsets are $C_{FB}$ = 1.781 and $C_{SD}$ = -0.583, in agreement with the values of the means. The phase difference is sufficiently different from $\pi$, which shows that the two oscillations do not cancel each other out in the TSI. Thus, the presence of this periodicity in the TSI wavelet diagram is also explained.

We also performed cross-wavelet analysis for sunspot darkening  and faculae brightening. The cross wavelet is used to more closely examine  the relation between the two time series \citep{Torrence1998}, in our case the faculae and sunspot contributions in TSI. It is based on the use of complex variables and allows us to calculate the cross-wavelet power.  The cross-wavelet power is displayed on the upper right panel of Figure ~\ref{fig:8}, which shows that the periods of 115 day, 140 day and 165 day simultaneously exist in sunspot darkening and faculae brightening as it is visually seen on wavelet spectra of both data. All the observed periods are collected in Table 1.

\subsection{Cycle 24}

TSI displays several peaks in cycle 24 (see lower panels in Figure ~\ref{fig:7}). Strong oscillation power is located at the periods of 170 days, 145 days and 115-120 days. Another weaker peak is seen around   210 days. The peak at 170 days is in the range of typical Rieger type periodicity, while other peaks are out of the range. Wavelet spectrum of full disk sunspot area data shows a strong, broad peak at 180-210 days, a weaker peak at 140 days and a very weak peak at 110 days. It is of importance to check the periodicities in the sunspot area of different hemispheres.
The lower panels of Figure~\ref{fig:10} show the wavelet spectra of hemispheric sunspot areas in cycle 24. The northern hemisphere shows the broad peak at the period of 190-210 days and weaker peak at 110 days, while the southern hemisphere displays the broad peak at 180-190 days and weaker peak at 120 days. There is a strong peak at 145 days, but almost outside the cone of influence.

The lower panel of Figure~\ref{fig:8} shows the wavelet spectra of the faculae brightening (left) and sunspot darkening (middle) from SATIRE reconstruction data for cycle 24. It can be seen that the strong periods of 170 days and 145 days in the TSI wavelet spectrum correspond to the sunspot darkening. The strong peak at 190 days in sunspot darkening is probably canceled by a corresponding period in faculae brightening when we look into the TSI. The weak peak at 210 days in the TSI is probably caused by faculae brightening. The oscillation at 115-120 days in TSI wavelet spectrum is probably generated by weak peaks at sunspot darkening.

%Following our analysis of the 115~d, 140~d and 165~d periodicity in cycle~23
We followed the same procedure as for cycle 23 with the 190~d periodicity and fit the function of Eq.~(\ref{eq:fit_funct}) to the facular brightening and the sunspot darkening in the periods of %\LEt{ Please write out these periods/time spans.}
2010-2016. The resulting amplitudes are $A_{FB}=0.1086$ and $A_{SD}=0.0857,$ while  the phase difference is $|\phi_{FB}-\phi_{SD}|=3.073$. We see that the oscillations are in anti-phase in both data, which leads to its cancelation in the TSI. The cross-wavelet power on the lower right panel of Figure ~\ref{fig:8} shows the simultaneous existence of 190-220 day oscillations in sunspot darkening and faculae brightening, as it is visually seen in their wavelet spectra.
All observed periods are collected in  Table 1.

In general, it can be seen that the sunspot contribution in the oscillation spectrum of TSI in the range of Rieger periodicity is stronger than the contribution of the faculae. Hence, the periods in TSI are mostly determined by the periods found in sunspot data.

%=====================
\section{Discussion and conclusions}
%=====================

Total solar irradiance can be considered as a light curve of the Sun as representative of a star for the purposes of this study. Therefore, a detailed investigation of temporal variation of TSI is a key point in stellar light curve analysis. Our knowledge of the physical processes in the solar atmosphere and interior is far more exhaustive than that of stellar case. It is of importance to know how these processes influence the behavior of TSI, so that we can could perform a reverse task in the stellar case to find out more about the processes in the stellar atmosphere and interior from the temporal behavior of their light curves.
We note that the contribution of the bright structures (faculae and networks) in the TSI increases at high latitudes, while the contribution of sunspots decreases \citep{Schatten1993}. Hence, TSI would be different when observed from different inclination angles. \citet{Nemec2020} also showed that inclination effects on TSI variability are significant. Therefore, the application of solar model to other stars with different inclination angles must be taken with caution.

Here we studied the dynamics of TSI over timescales of Rieger-type periodicity in the solar cycles 23 and 24 using the wavelet transform. There are several findings that could be of importance in  subsequent studies of stellar light curves.

\begin{figure*}
\includegraphics[width=17cm]{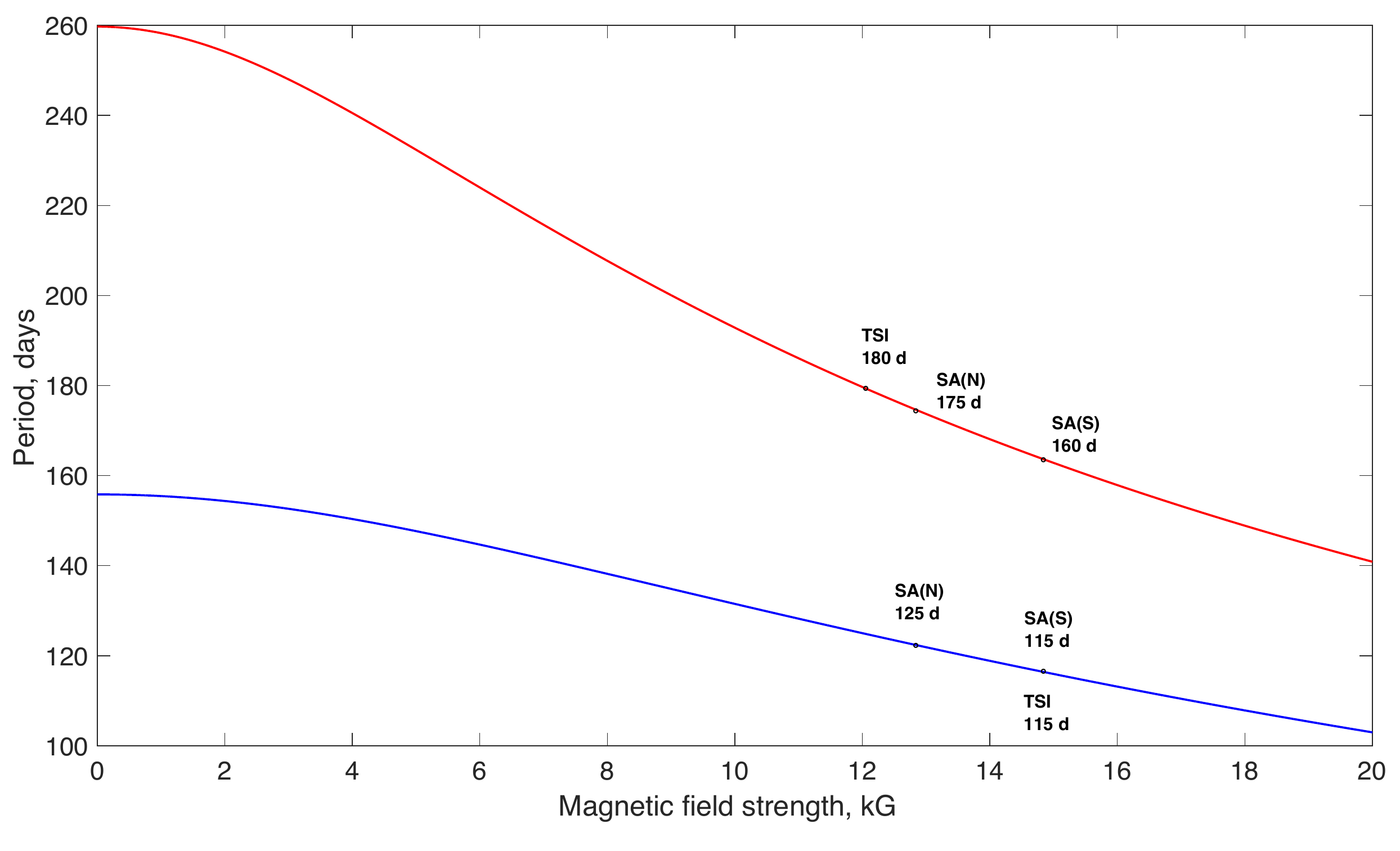}
\caption{Period versus dynamo magnetic field strength of the magneto-Rossby wave $m=1$ spherical harmonics with $n=4$ (red curve) and $n=3$ (blue curve) plotted according to Eq. (43) from \citet{Gachechiladze2019}. The equatorial angular velocity and plasma density in the dynamo layer are assumed to be $\Omega_0=2.8 \cdot 10^{-6}$ s$^{-1}$ and $\rho=0.2$ g cm$^{-3}$, respectively. The observed periods in different data sets of the cycle 23 are marked by dots on the curves. }
 \label{fig:12}
\end{figure*}

The wavelet spectrum of TSI gives two pronounced peaks at the periods of $\sim$ 180 days and 115 days in cycle 23. The period of 180 days also appears in full-disk sunspot data (see the right upper panel of Figure~\ref{fig:7} and the middle upper panel of Figure ~\ref{fig:8}), but not in the faculae data (the left upper panel of Figure ~\ref{fig:8}). On the other hand, the hemispheric sunspot wavelet spectra show that the period is connected to the northern hemisphere. The southern hemisphere displays around 160-day periodicity, which is absent in the TSI. The period of 115 days also appears in full-disk sunspot data, and contrary to the 180 days, it is probably connected to the southern hemisphere (see Figure~\ref{fig:10}). Another weak peak in the power spectrum of TSI is around 145 days, which is also seen in full-disk sunspot spectrum. Therefore, it is clear that the variations in TSI seem to be determined by sunspot darkening rather than faculae brightening. There are two peaks in sunspot data of the cycle 23 at 165 days and 140 days, which are entirely absent in the TSI spectrum. To understand the reason for the disappearance, we should carefully look into the faculae and sunspot contributions (Figure~\ref{fig:8}).  Cross-wavelet analysis shows that both periods appear in the faculae and sunspot data simultaneously; therefore, faculae brightening and sunspot darkening may cancel each other out in the TSI. Indeed, a fitting of the sinusoidal function to the data shows that the 140-day oscillations are in an anti-phase (Figure~\ref{fig:11}); therefore, the oscillations have almost no contribution. The oscillations with the period of 165 days in both sets of data are not in the anti-phase, but it is likely that they are not seen in the TSI because of their low power. On the other hand, the period of 180 days has no counterpart in the faculae data; therefore, it can be seen in the TSI spectrum. Another period of 115 days in the TSI spectrum is seen in both the sunspot darkening and faculae brightening. However, the period reaches peaks at different times; therefore it is not cancelled out in the TSI spectrum.

%Cycle 24 shows basically similar properties as cycle 23, however with slightly different periods.
The TSI displays the strong periodicity at 170 days and 145 days in cycle 24, while weaker periods are seen around 210 days and 110-115 days. All these periods are probably connected to sunspot darkening rather than faculae brightening. The period of 190 days seen in the sunspot data is probably canceled in the TSI spectrum due to the same periodicity in the faculae brightening (see lower panel of Figure~\ref{fig:8}).

It is very important to understand the physical mechanism, which is responsible for the periodicity in TSI. Rieger-type periodicity is supposed to be caused by the instability of magnetic Rossby waves in the solar dynamo layer due to the differential rotation and toroidal magnetic field \citep{zaqarashvili2010}. Rossby wave dispersion relation in the simplest nonmagnetic case is:
\begin{equation}
\label{disp}
\omega=-\frac{2 m \Omega_0}{n(n+1)},
\end{equation}
where $\omega$ is the wave frequency and $\Omega_0$ is the equatorial angular velocity, while $m$ and $n$ are angular order and degree of corresponding spherical harmonics ($m\leq n$).
Therefore, observed periods may appear as harmonics of the rotation period. In the nonmagnetic case, $m=1$ and $n=3$ harmonics gives the period of 156 days for the rotation rate of 26 days. This is the same value found by \citet{Rieger1984}. However, the observed periods are significantly different from the simple harmonics of the solar rotation period. This is because that the toroidal magnetic field of the dynamo layer, which is supposed to operate below the convection zone, modifies the dispersion relation, so that the period of magnetic Rossby waves depends on the field strength and its latitudinal structure \citep{Zaqarashvili2021}. Therefore, we can estimate the toroidal magnetic field in the dynamo layer using the observed periods and the dispersion relation of magnetic Rossby waves. For the latitudinal structure of the magnetic field, we used the formula $B_{tor}=B_0 \sin \theta \cos \theta$, where $\theta$ is the latitude, which resembles the solar dynamo field (it has different signs in different hemispheres and reaches maximum at mid latitudes). Figure ~\ref{fig:12} shows the period vs magnetic field strength for spherical harmonics of magneto-Rossby waves, with $m=1$ and $n=3,4$ according to the Eq. (43) from \citet{Gachechiladze2019}. It was shown that only $m=1$ harmonics are unstable below the convection zone \citep{zaqarashvili2010}, therefore, we consider only these harmonics, but with different $n$. We see that the magnetic field with a strength of 12 kG yields the periods of 180 days and 125 days for the harmonics of $n=4$ and $n=3$, respectively. These values fairly correspond to the periods found in TSI of the solar cycle 23 (180 days and 115 days). However, hemispheric sunspot areas display the periods of 175 days and 125 days in the northern hemisphere and the periods of 160 days and 115 days in the southern hemisphere. According to Figure ~\ref{fig:12}, the periods correspond to 13 kG and 15 kG magnetic field strength in the northern and southern hemispheres, respectively. Therefore, if the Rossby wave scenario is compatible with the Rieger-type periodicities, then we can use the TSI oscillations to estimate the toroidal magnetic field strength. As the dynamo theory is believed to be responsible for generation of toroidal magnetic field in the solar interior, then the estimated field strength clearly belongs to the dynamo-generated field.
In a similar way, the observed Rieger-type periodicity in the light curves of solar-like stars and the dispersion relation of magneto-Rossby waves may allow us to estimate the magnetic field strength in the stellar dynamo layers. This results in a rather rough estimation, but by using many stars with different rotation periods (i.e., at different stages of evolution), this could provide information on a general feature of magnetic field change through stellar age.

\begin{acknowledgements}

EG was supported by Shota Rustaveli National Science Foundation of Georgia (SRNSFG) [grant number N04/46, project number N93569] and by Volkswagen Foundation in the framework of the joint project "Structured Education - Quality Assurance - Freedom to Think". TVZ was supported by the Austrian Fonds zur F{\"o}rderung der Wissenschaftlichen Forschung (FWF) project P30695-N27. RO acknowledges support from grant AYA2017-85465-P(MINECO/AEI/FEDER, UE). TR acknowledges support from the European Research Council (ERC) under the European Union's Horizon 2020 research and innovation program (grant agreement No. 715947). This paper resulted from discussions at workshops of ISSI (International Space Science Institute) team (ID 389) "Rossby waves in astrophysics" organised in Bern (Switzerland). We thank C. B. Markwardt for making available at \url{http://purl.com/net/mpfit} the IDL implementation of the MPFIT  library. V6.4 of the VIRGO Experiment on the cooperative ESA/NASA Mission SoHO from VIRGO Team through PMOD/WRC, Davos, Switzerland

\end{acknowledgements}

\end{document}